# Analysis of the Emission and Morphology of the Pulsar Wind Nebula Candidate HAWC J2031+415

R. Alfaro[1], C. Alvarez[2], J. C. Arteaga-Velázquez[3], D. Avila Rojas[1], H. A. Ayala Solares[4], R. Babu[5], E. Belmont-Moreno[1], K. S. Caballero-Mora[2], T. Capistrán[6], A. Carramiñana[7], S. Casanova[8], U. Cotti[3], J. Cotzomi[9], S. Coutiño de León[10], E. De la Fuente[11], C. de León[3], D. Depaoli[12], N. Di Lalla[13], R. Diaz Hernandez[7], B. L. Dingus[14], M. A. DuVernois[10], J. C. Díaz-Vélez[10], K. Engel[15], T. Ergin[5], C. Espinoza[1], K. L. Fan[15], N. Fraija[6], J. A. García-González[16], M. M. González[6], J. A. Goodman[15], S. Groetsch[17], J. P. Harding[14], S. Hernández-Cadena[18], I. Herzog[5], D. Huang[15], F. Hueyotl-Zahuantitla[2], P. Hüntemeyer[17], A. Iriarte[6], S. Kaufmann[19], J. Lee[20], H. León Vargas[1], A. L. Longinotti[6], G. Luis-Raya[19], K. Malone[14], J. Martínez-Castro[21], J. A. Matthews[22], P. Miranda-Romagnoli[23], J. A. Montes[6], E. Moreno[9], M. Mostafá[24], M. Najafi[17], L. Nellen[25], M. Newbold[26], M. U. Nisa[5], R. Noriega-Papaqui[23], Y. Pérez Araujo[1], E. G. Pérez-Pérez[19], C. D. Rho[27], D. Rosa-González[7], E. Ruiz-Velasco[12], H. Salazar[9], A. Sandoval[1], D. Salazar-Gallegos[5], M. Schneider[15], J. Serna-Franco[1], A. J. Smith[15], Y. Son[20], R. W. Springer[26], O. Tibolla[19], K. Tollefson[5], I. Torres[7], R. Torres-Escobedo[18], R. Turner[17], F. Ureña-Mena[7], E. Varela[9], L. Villaseñor[9], X. Wang[17], Zhen Wang[15], I. J. Watson[20], S. Yu[4], S. Yun-Cárcamo[15], and H. Zhou[18]
(THE HAWC COLLABORATION)

[1] Instituto de Física, Universidad Nacional Autónoma de México, Ciudad de México, Mexico
[2] Universidad Autónoma de Chiapas, Tuxtla Gutiérrez, Chiapas, Mexico
[3] Universidad Michoacana de San Nicolás de Hidalgo, Morelia, Mexico
[4] Department of Physics, Pennsylvania State University, University Park, PA, USA
[5] Department of Physics and Astronomy, Michigan State University, East Lansing, MI, USA; herzogia@msu.edu
[6] Instituto de Astronomía, Universidad Nacional Autónoma de México, Ciudad de México, Mexico
[7] Instituto Nacional de Astrofísica, Óptica y Electrónica, Puebla, Mexico
[8] Instytut Fizyki Jadrowej im Henryka Niewodniczanskiego Polskiej Akademii Nauk, IFJ-PAN, Krakow, Poland
[9] Facultad de Ciencias Físico Matemáticas, Benemérita Universidad Autónoma de Puebla, Puebla, Mexico
[10] Department of Physics, University of Wisconsin-Madison, Madison, WI, USA
[11] Departamento de Física, Centro Universitario de Ciencias Exactase Ingenierias, Universidad de Guadalajara, Guadalajara, Mexico
[12] Max-Planck Institute for Nuclear Physics, 69117 Heidelberg, Germany
[13] Department of Physics, Stanford University, Stanford, CA 94305-4060, USA
[14] Los Alamos National Laboratory, Los Alamos, NM, USA
[15] Department of Physics, University of Maryland, College Park, MD, USA
[16] Tecnologico de Monterrey, Escuela de Ingeniería y Ciencias, Ave. Eugenio Garza Sada 2501, Monterrey, N.L. 64849, Mexico
[17] Department of Physics, Michigan Technological University, Houghton, MI, USA
[18] Tsung-Dao Lee Institute & School of Physics and Astronomy, Shanghai Jiao Tong University, Shanghai, People's Republic of China
[19] Universidad Politecnica de Pachuca, Pachuca, Hgo, Mexico
[20] University of Seoul, Seoul, Republic of Korea
[21] Centro de Investigación en Computaci'on, Instituto Politécnico Nacional, México City, Mexico
[22] Department of Physics and Astronomy, University of New Mexico, Albuquerque, NM, USA
[23] Universidad Autónoma del Estado de Hidalgo, Pachuca, Mexico
[24] Department of Physics, Temple University, Philadelphia, PA, USA
[25] Instituto de Ciencias Nucleares, Universidad Nacional Autónoma de México, Ciudad de México, Mexico
[26] Department of Physics and Astronomy, University of Utah, Salt Lake City, UT, USA
[27] Department of Physics, Sungkyunkwan University, Suwon 16419, Republic of Korea
Received 2024 June 17; revised 2024 August 29; accepted 2024 September 6; published 2024 November 4

## Abstract

The first TeV $\gamma$-ray source with no lower energy counterparts, TeV J2032+4130, was discovered by HEGRA. It appears in the third HAWC catalog as 3HWC J2031+415 and it is a bright TeV $\gamma$-ray source whose emission has previously been resolved as two sources: HAWC J2031+415 and HAWC J2030+409. While HAWC J2030+409 has since been associated with the Fermi Large Area Telescope Cygnus Cocoon, no such association for HAWC J2031+415 has yet been found. In this work, we investigate the spectrum and energy-dependent morphology of HAWC J2031+415. We associate HAWC J2031+415 with a $\gamma$-ray binary system containing the pulsar PSR J2032+4127 and its companion MT91 213. We study HAWC data to observe their periastron in 2017. Additionally, we perform a combined multiwavelength analysis using radio, X-ray, and $\gamma$-ray emission. We conclude that HAWC J2031+415 and, by extension, TeV J2032+4130 are most probably a pulsar wind nebula powered by PSR J2032+4127.

*Unified Astronomy Thesaurus concepts:* Gamma-ray astronomy (628); Pulsar wind nebulae (2215); Binary pulsars (153); Gamma-ray sources (633)

## 1. Introduction

First observed in 2005 by the High Energy Gamma Ray Astronomy (HEGRA) experiment, TeV J2032+4130 was the first very high energy (>100 GeV) γ-ray source in the TeV range with no lower energy counterpart (F. Aharonian et al. 2005). TeV J2032+4130 is located in the Fermi Large Area Telescope (LAT) Cygnus Cocoon region, a large extended source of GeV γ-ray emission that contains the Cygnus OB-2 star cluster (M. Ackermann et al. 2011). The extent of the original HEGRA detection was 0°.11 and was comparatively dim to HEGRA's detection of the Crab above 1 TeV at 5% of the Crab's flux (G. Rowell et al. 2003; F. Aharonian et al. 2005).

Follow-up studies by the X-ray observatories Suzaku, Chandra, and XMM-Newton revealed significant diffuse nonthermal X-ray emission coincident with TeV J2032+4130 (D. Horns et al. 2007; H. Murakami et al. 2011). In H. Murakami et al. (2011), they revealed two substructures, one of which was coincident with the pulsar PSR J2032+4127. In addition, they also observed a large diffuse excess measured across TeV J2032+4130's extent (H. Murakami et al. 2011). XMM-Newton's detection is roughly the same size as TeV J2032+4130, though no substructures were found (D. Horns et al. 2007). Both measurements had fluxes significantly lower than that of the γ-ray source. Two hypotheses were proposed for the emission: hadronic, driven by pion decay, and leptonic, produced via a combination of synchrotron (SYN) and inverse Compton (IC) scattering. Both hypotheses are considered in this analysis and are discussed in Section 6.

Radio observations made using the Very Large Array (VLA) revealed a large number of radio sources in the direction of TeV J2032+4130's center of gravity (COG). One of these sources was characterized by faint, nonthermal emission in roughly a half-circle around the COG with a total area of ∼27 arcmin² (J. M. Paredes et al. 2006; J. Marti et al. 2007). The region had an estimated energy content of $6 \times 10^{45}$ erg and seemed to indicate an efficient injector of nonthermal particles. Additionally, the semicircular shape of the emission region seems to indicate an old supernova shell and may be the radio counterpart of TeV J2032+4130.

Later TeV observations by the Very Energetic Radiation Imaging Telescope Array System (VERITAS) in 2014 (E. Aliu et al. 2014) and 2018 (A. U. Abeysekara et al. 2018a) found emission that corresponded to an asymmetric Gaussian within the energy range 0.5–50 TeV. In both E. Aliu et al. (2014) and A. U. Abeysekara et al. (2018a), they hypothesize that the emission is from a pulsar wind nebula (PWN) and whose source is PSR J2032+4127. Furthermore, while they do not observe one, they predicted a cutoff near 10 TeV.

In the second (A. U. Abeysekara et al. 2017b) and third (A. Albert et al. 2020) catalogs published by the High-Altitude Water Cherenkov (HAWC) observatory, sources 2HWC and 3HWC J2031+415 were detected coincident with TeV 2032 +4130. A follow-up dedicated analysis resolved two sources: HAWC J2031+415, which was associated with the probable PWN, and HAWC J2030+409, believed to be the TeV extension of the Fermi-LAT Cygnus Cocoon (A. U. Abeysekara et al. 2021a). Though that analysis focused on the Cygnus Cocoon (henceforth referred to as the Cocoon), it was found that HAWC J2031+415 had an extension of 0°.27 and a power law (PL) with an exponential cutoff spectral model with a cutoff on the order of >10 TeV and is consistent with VERITAS' observations (E. Aliu et al. 2014; A. U. Abeysekara et al. 2021a).

As asserted by VERITAS, PSR J2032+4127 is most likely the power source for the PWN. PSR J2032+4127 is an unique pulsar to power a PWN. First, it is old at an estimated characteristic age of ∼200 kyr and has an estimated spin-down luminosity $\dot{E} = 1.5 \times 10^{35}$ erg s$^{-1}$. Current estimates have moved it from 3.8 kpc in H. Murakami et al. (2011) to $1.33 \pm 0.06$ kpc in the most recent pulsar catalog published by the Australian Telescope National Facility (R. N. Manchester et al. 2005). This places it inside the Cocoon, which has a distance of ∼1.4 kpc (M. Ackermann et al. 2011; A. U. Abeysekara et al. 2021a). Additionally, it is a long period binary with the ∼15 $M_\odot$ star MT91 213 and has an orbital period of 50 yr (A. G. Lyne et al. 2015). This makes the system unique, as TeV binary γ-ray pulsar systems are rare. While not originally associated with observed X-ray emission (A. G. Lyne et al. 2015), the pulsar is now believed to be responsible for it (E. Aliu et al. 2014) and will be considered for the multiwavelength analysis presented in Section 6. Additionally, in 2017 November, it performed its periastron with MT91 213 and flaring in both γ- and X-rays was detected (A. U. Abeysekara et al. 2018b). This is investigated using HAWC data.

In this paper, we further study the probable PWN HAWC J2031+415. In Section 2, we introduce the HAWC observatory. Section 3 discusses the analysis pipeline that we use to describe HAWC J2031+415. Section 4 explores the energy-dependent morphology of HAWC J2031+415. Section 5 discusses the results of HAWC's analysis of the 2017 periastron. Section 6 incorporates data from other observatories to perform a multiwavelength analysis on this source. Section 7 summarizes our conclusions.

## 2. HAWC

The HAWC observatory is located on the extinct volcano Sierra Negra in Mexico. The detector is at an altitude of 4100 m with a main array of 300 water Cherenkov detectors, each containing four photomultiplier tubes (PMTs), and covers an area of 22,000 m². It is sensitive to γ-rays in the range of ≃0.3 to ≃300 TeV. Additional details can be found in A. U. Abeysekara et al. (2023). For this analysis, 2400 days of data, approximately 1000 days more data than in A. U. Abeysekara et al. (2021a), are used. A neural network (NN) algorithm is used to reconstruct the energy of each recorded γ-ray. This network utilizes three general inputs: the amount of energy deposited, the amount the shower is contained in the detector's footprint, and the attenuation degree caused by the atmosphere. Once the energy is reconstructed, it is then binned into a 2D binning scheme that takes both the reconstructed energy and the fraction of PMTs triggered during the extensive air shower (A. Albert et al. 2024a). Along with the additional data, this data reconstruction process uses superior background-rejection algorithms used in A. U. Abeysekara et al. (2021a).

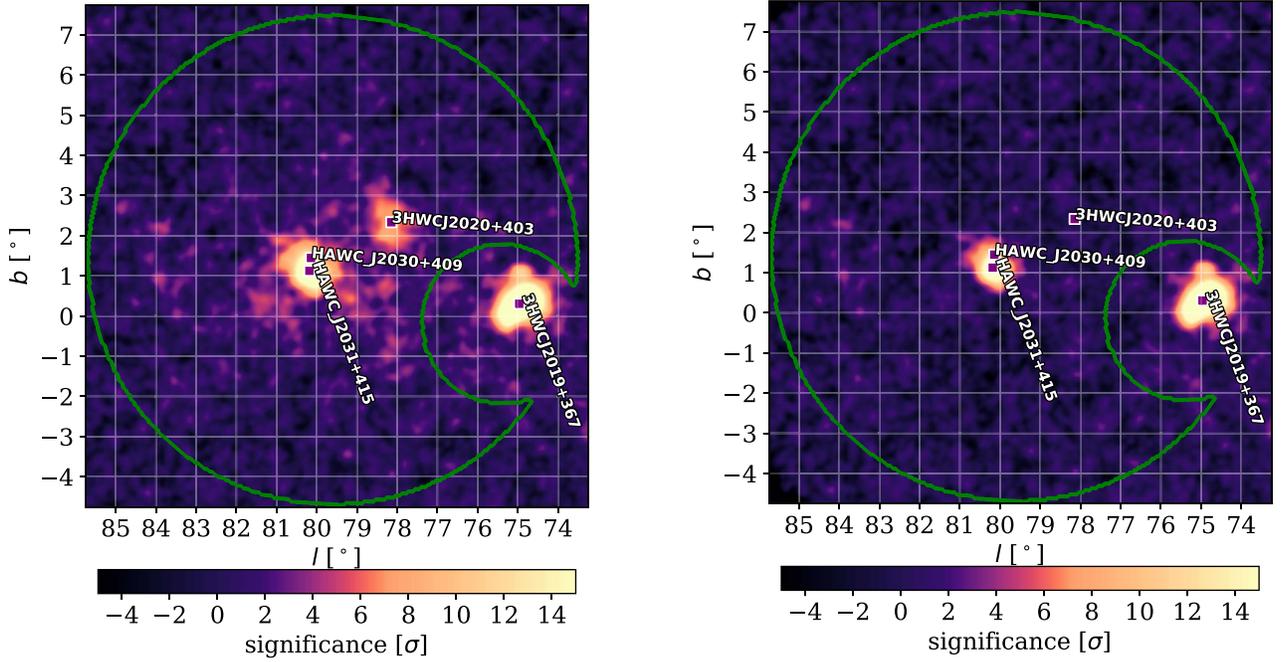

**Figure 1.** Left: a significance map of the ROI (green contour) with the source associations found in Section 3. A mask is placed on 3HWC J2019+367 to avoid contamination from its emission for this analysis. Right: HAWC J2031+415's emission is shown after contributions from HAWC J2030+409 and 3HWC J2020+403 were subtracted from the data map.

## 3. Source Search and Spectral Fitting

### 3.1. Region of Interest Considered

The data described above is subdivided into a region of interest (ROI). The ROI is a 6° circular region centered on $(l = 78°.9, b = 1°.6)$ with a mask on 3HWC J2019+367. This is shown in Figure 1. As done in A. U. Abeysekara et al. (2021a), the mask on 3HWC J2019+367 is used to prevent potential contamination caused by the brightest source in the Cocoon region.

### 3.2. Method

To fit the $\gamma$-ray data, we used both the Multi-Mission Maximum Likelihood framework (G. Vianello 1938)[28] and the HAWC Accelerated Likelihood[29] (A. U. Abeysekara et al. 2021b) plugin. This implementation allows extensive multi-source fitting for complex regions. The framework considers a test statistic (TS) that evaluates the statistical significance of a given model with a given number of free parameters. The TS is used to compare an alternative hypothesis with a null hypothesis. It is defined as

$$\mathrm{TS} = 2\ln\left(\frac{L_{\mathrm{alt}}}{L_{\mathrm{null}}}\right). \quad (1)$$

If two alternate nested hypotheses are compared, $\Delta \mathrm{TS} = \mathrm{TS}_2 - \mathrm{TS}_1 = 2\ln(L_2/L_1)$ can be used to determine which model is preferred (A. U. Abeysekara et al. 2017a). If the difference in free parameters between the models is 1, then Wilks' theorem can be used to give a pretrial significance that follows $\sigma = \sqrt{\mathrm{TS}}$ (S. S. Wilks 1938).

To fully model the emission in the ROI, we performed a source search method similar to that of the Fermi-LAT extended source catalog (M. Ackermann et al. 2017). All models considered are from the Astromodels[30] Python package. This search method can be broken into three broad sections: a point-source (PS) search, an extended (EXT) text, and then a spectral test. The process is defined by the following steps.

1. First, contributions from both unresolved source emission and the diffuse background emission is modeled using the unresolved radiation model (URM). This model uses a 2D Gaussian template centered at 0° along the Galactic plane. The spectral model is defined as a PL defined by

$$\frac{dN}{dE} = N_o\left(\frac{E}{E_p}\right)^{-\alpha}, \quad (2)$$

where $N_o$ and $\alpha$ are the flux normalization and index for the source, respectively. $E_p$ is the fixed pivot energy for the source and is selected to minimize the correlation between the flux normalization and index. As done in A. Albert et al. (2024b), a spectral index of 2.7 and a pivot of 7 TeV is assumed while $N_o$ is free to float with the fit. After the fit, the TS is calculated and the ROI is searched for any remaining positive excess.

2. A floating PS is added to the URM model (1 PS + URM) at the pixel of highest remaining emission. The ROI is then refitted with this new model (PS + URM), the TS is calculated, and then compared to the URM only model. If the $\Delta \mathrm{TS} > 25$, then the 1 PS + URM is the preferred model and is kept as the new preferred model.

3. Another floating PS is then added to the model at the next pixel of highest significance and the process is repeated. This continues until adding a PS does not give a $\Delta \mathrm{TS} > 25$, in which case the PS search is completed.

---
[28] https://github.com/threeML/threeML
[29] https://github.com/threeML/hawc_hal
[30] https://github.com/threeML/astromodels

The final PS is excluded from this last model as it failed to cross the necessary TS threshold.

4. All sources must initially be detected as PSs, regardless if they are extended or not. An EXT is a 2D symmetric Gaussian model defined by

$$\text{EXT} = \left(\frac{180}{\pi}\right)^2 \frac{1}{2\pi\sigma^2} \exp\left(-\frac{\vec{\theta}^2}{2\sigma^2}\right), \quad (3)$$

where the $\left(\frac{180}{\pi}\right)^2$ factor converts degrees to radians and the $\vec{\theta}$ and $\sigma$ parameters are the location and radius of the Gaussian, respectively. The brightest PS is converted to an EXT and the region is refit with the location of all sources fixed but the index and flux normalization free. There are two possible outcomes:

(a) if $\Delta\text{TS} > 25$, then the extension is preferred and kept. After the fit, all sources that have a $\text{TS} < 25$ are dropped from the model; and
(b) if $\Delta\text{TS} < 25$, the extension is rejected and the model reverts to the preextension test.

5. Regardless of the previous EXT test, the next brightest PS is then tested. This process repeats until all sources have been tested. Once the extension test is completed, all the remaining sources in the ROI are refit with free locations.

6. In a similar fashion, the spectrum of each source is now tested. There are three models considered: the PL model in Equation (2), a power law with an exponential cutoff (PLC), and a log-parabola (LP) model. The latter two are given by Equations (4) and (5), respectively

$$\frac{dN}{dE} = N_o \left(\frac{E}{E_p}\right)^{-\alpha} \exp\left(-\frac{E}{E_c}\right), \quad (4)$$

$$\frac{dN}{dE} = N_o \left(\frac{E}{E_p}\right)^{-\alpha - \beta \ln(E/E_p)}. \quad (5)$$

The extra terms $E_c$ and $\beta$ are the cutoff energy and curvature of the spectrum, respectively.

7. To compare the spectral models, the Akaike information criterion (AIC) is used and is given by (H. Akaike 1974)

$$\text{AIC} = 2k - 2\ln(\hat{L}). \quad (6)$$

The AIC adds a term that penalizes more complex models given by the $k$ number of parameters for a given maximized likelihood $\hat{L}$. As outlined in D. Anderson & K. Burnham (2004), the lower the AIC, the more preferred a model is. A $\Delta$AIC between 0 and 2 indicates both models fit well and the simpler should be considered and, for $\Delta$AIC > 3, the higher AIC model becomes decreasingly likely.

### 3.3. Results

Once the source search process has been completed, the final source model contains the following: URM and three extended sources: HAWC J2031+415, HAWC J2030+409, and 3HWC J2020+403 and is in agreement with A. U. Abeysekara et al. (2021a). HAWC J2030+409 is associated with the Fermi-LAT Cocoon (M. Ackermann et al. 2017). HAWC J2031+415 is near coincident with TeV J2032+4130 and is assumed to be its higher energy counterpart. 3HWC J2020+403 is coincident with the supernova γ Cygni and has been also observed by Fermi-LAT (S. Abdollahi et al. 2020).

The results from the fitting process are given in Table 1 and the preferred spectral models for the sources being as follows: HAWC J2031+415 is a PLC, HAWC J2030+409 is an LP, and 3HWC J2020+403 is a PL. A brief comparison to the previously published work is discussed in Section 3.4.

The systematic uncertainties given in Table 1 are found by performing a series of fits with detector response files that describe different detector configurations. Further details are given in A. U. Abeysekara et al. (2019) and A. Albert et al. (2024a) and are briefly summarized here. These response files are generated assuming different PMT response to showers (efficiency over time, response, etc.) and are then compared to HAWC's standard response file. The fitting process is then repeated with these new response files, the difference between the new fit values and those in Table 1 are found, and then are added in quadrature to produce the total systematic uncertainties.

We determine the energy range of each source using the following procedure in A. U. Abeysekara et al. (2017c). Each spectral model is independently multiplied by a step function that models an abrupt cutoff in the spectrum. The only free parameters are the cutoff values; all other parameters are fixed at their best-fit values. These values float until the TS of significance of the sources drops by $1\sigma$. This gives the $1\sigma$ energy limits for the sources as follows in units of TeV: 0.4–151 for HAWC J2031+415, 0.5–250 for HAWC J2030+409, and 0.26–100 for 3HWC J2020+403.

With the multisource fitting process complete, we then isolate the emission of HAWC J2031+415 by subtracting out the modeled emission of the URM, HAWC J2030+409, and 3HWC J2020+403 from Figure 1 (left). The result is shown in Figure 1 (right).

Figure 2 shows the spectral energy distribution (SED) of HAWC J2031+415 compared to other selected observations. These observations represent the most current detections of TeV J2032+4130 from their respective observatories. It can be seen that HAWC's observation is in conflict with all other observations. This can be explained by HAWC J2031+415's much larger extension compared to the previous studies. For example, in HEGRA's initial discovery, TeV J2032+4130 had an extent of 0°.11 compared to this work's 0°.26, thus leading to a much larger flux. Of special note is the scaling done to VERITAS's measurement. We follow the procedure outlined in A. Albert et al. (2021) and is briefly described below.

The spectrum reported by VERITAS was found from a smaller region than the full observed region presented in A. U. Abeysekara et al. (2018a). VER J2031+415's morphology is described as an asymmetric Gaussian with extents $0°.15 \pm .03°$ and $0°.07 \pm 0°.01$ for the semimajor and semiminor axes, respectively, with a $63°$ rotation to the northwest. By contrast, the flux calculation uses a circular region with radius 0°.23 centered on VER J2031+415. This method is different to what is used in this analysis, where the flux calculation is computed concurrently with the morphological fit. As such, the flux measurements of VERITAS and HAWC may be systematically offset for larger extended sources like HAWC J2031 +415. To account for this offset, the flux reported by VERITAS is scaled by assuming a larger integration region. This gives a scaling factor of 1.49 and is used in Figure 2.

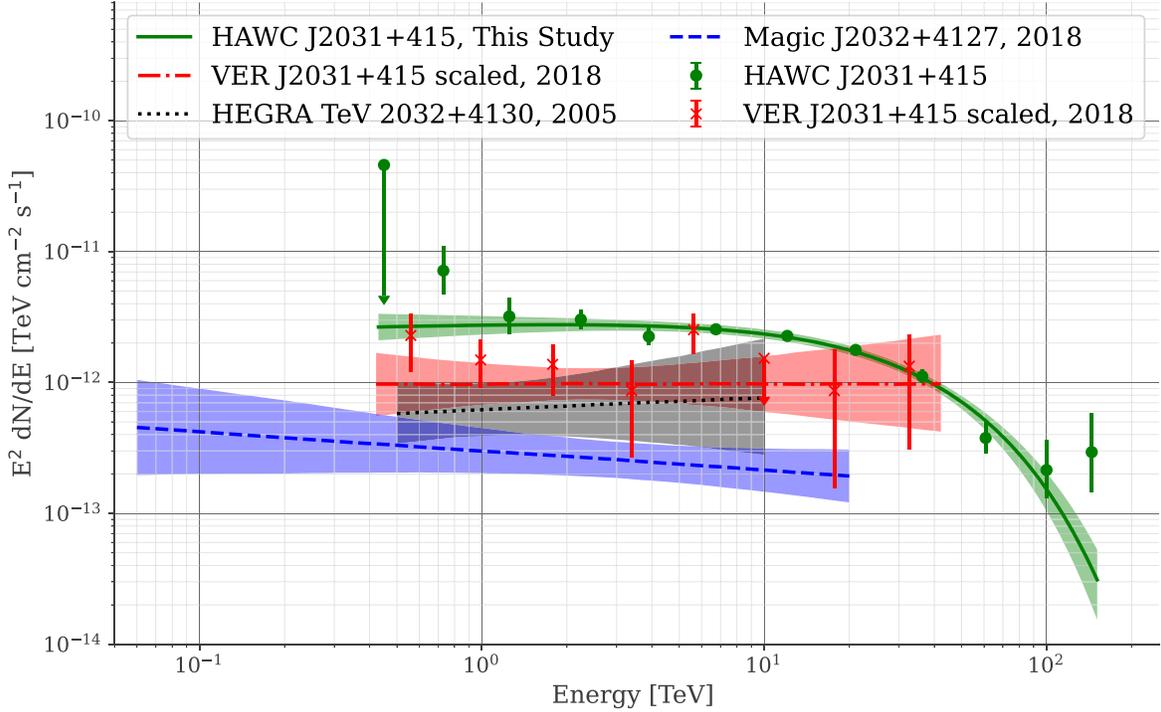

**Figure 2.** SED of HAWC J2031+415. The other observations are from F. Aharonian et al. (2005), J. Albert et al. (2008), and A. U. Abeysekara et al. (2018b, 2018a), respectively, and were selected as the most current independent observations available. Additionally, all uncertainties are statistical only.

**Table 1**
Fit Results from the Systematic Source Search

| Source Name | Spectral Parameters | Morphology |
|---|---|---|
| HAWC J2031+415 | $\phi_{4.9\,\text{TeV}} = 1.29^{+0.14+0.15}_{-0.12-0.25} \times 10^{-13}$<br>$\alpha = 1.94^{+0.10+0.10}_{-0.10-0.19}$<br>$E_c = 32^{+7+5}_{-7-4}$ | $\sigma = 0.255^{+0.016+0.015}_{-0.016-0.019}$<br>R.A. $= 307.92^{+0.02+0.01}_{-0.02-0.01}$<br>Decl. $= 41.48^{+0.02+0.01}_{-0.02-0.01}$ |
| HAWC J2030+409 | $\phi_{4.2\,\text{TeV}} = 1.1^{+0.12+0.20}_{-0.11-0.09} \times 10^{-12}$<br>$\alpha = 2.59^{+0.07+0.08}_{-0.07-0.18}$<br>$\beta = 0.11^{+0.04+0.02}_{-0.04-0.05}$ | $\sigma = 2.50^{+0.26+0.25}_{-0.26-0.48}$<br>R.A. $= 307.54^{+0.22+0.10}_{-0.22-0.24}$<br>Decl. $= 41.64^{+0.24+0.10}_{-0.24-0.31}$ |
| 3HWC J2020+403 | $\phi_{1.1\,\text{TeV}} = 4.3^{+0.8+0.6}_{-0.7-0.3} \times 10^{-12}$<br>$\alpha = 2.91^{+0.07+0.04}_{-0.07-0.09}$ | $\sigma = 0.36^{+0.05+0.03}_{-0.05-0.02}$<br>R.A. $= 305.05^{+0.07+0.05}_{-0.07-0.03}$<br>Decl. $= 40.52^{+0.05+0.03}_{-0.05-0.03}$ |

**Note.** The first uncertainty listed is statistical and the second is systematic. The units are as follows: $\phi_{E_p}$ is the flux normalization with units (TeV cm$^2$ s)$^{-1}$; R.A., decl., and $\sigma$ are given in degrees; and $E_c$ has units of TeV. $E_p$ was found for each source independently.

### 3.4. Comparison to Previous Work

This work found identical (within uncertainties) morphological and spectral models for HAWC J2031+415 and spectral model for 3HWC J2020+403 as in A. U. Abeysekara et al. (2021a) but there is tension with HAWC J2030+409's spectral model. In A. U. Abeysekara et al. (2021a), the preferred model was a PL with $\phi_{4.2\,\text{TeV}} = 9.3^{+0.9+0.93}_{-0.8-1.23} \times 10^{-13}$ TeV (cm$^2$ s)$^{-1}$ and $\gamma = 2.64^{+0.05+0.09}_{-0.05-0.03}$ while this work found an LP model to be strongly preferred with $\Delta$AIC = 90. This is explained by the superior background rejection utilized with the newer data set that reveals a curvature at the highest energies. The spectral and morphological fits for HAWC J2031+415 and 3HWC J2020+403 are comparable to A. U. Abeysekara et al. (2021a). One additional note is the recently published LHAASO result of the Cygnus region (LHAASO Collaboration 2024) where they find a PL spectral model for HAWC J2030+409. An in-depth comparison is beyond the scope of this work, but the two models are compatible within the systematic uncertainties.

For 3HWC J2020+403, a minor difference in morphology compared to H. Fleischhack (2019) and A. U. Abeysekara et al. (2021a) was found. In H. Fleischhack (2019), its model was found to be a disk rather than a symmetric 2D Gaussian. For a disk model, the emitted flux is held constant over a fixed radius rather than decreasing radially with a 2D Gaussian. We tested this by creating a disk model with a fixed radius of 0°.63 from H. Fleischhack (2019) and the whole model was refitted. The result was a negligible difference in TS ($\Delta$TS < 1) and, while the Gaussian model is used for this analysis, a dedicated work on 3HWC J2020+403 is need to determine its true morphology.

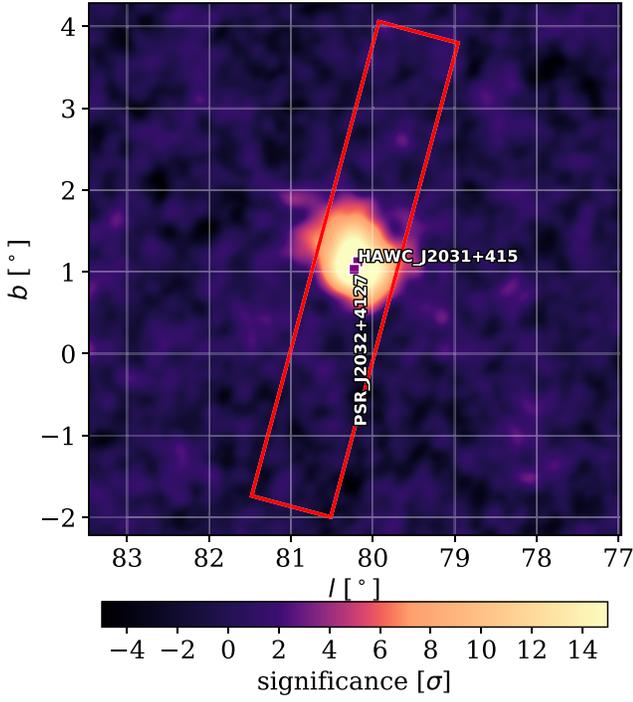

**Figure 3.** The significance map used for the energy morphology study. The rectangle highlights the longitudinal profile region used for the energy-dependent morphology study.

## 4. Energy-dependent Morphology Study

### 4.1. Methodology

In order to study any possible energy-dependent morphology of HAWC J2031+415, we utilized the method described in V. Joshi (2019) and A. Albert et al. (2021). This method uses the longitudinal profiles of discrete energy bands over the source to count the number of excess events observed. Six energy bands are selected in TeV units: 0.3–1.0, 1.0–3.2, 3.2–10, 10–32, 32–100, and 100–316. Each energy band consists of two NN energy bins, similar to V. Joshi (2019) and whose specific energy definitions are defined in A. Albert et al. (2024a).

The longitudinal profile region is defined as a rectangle of dimensions $6°$ long by $1°$ centered at the pulsar's location with HAWC J2030+409, 3HWC J2020+403, and the URM subtracted out. Furthermore, the rectangle is rotated by $15°$ to lie on the line connecting HAWC J2031+415's centroid and PSR J2032+4127 in Galactic coordinates. This is to determine whether the observed emission trends toward the pulsar's location with changing energy. This can be seen in Figure 3.

To determine the true size of the extended emission in each energy band, the following procedure is used. First, the rectangular region is divided into 50 bins, each with a width of $0°.12$. Then the excess counts of each bin are summed and plotted. This is shown by the data points in Figure 4. To measure the intrinsic width of the extension, a 1D Gaussian is fit to each band, as indicated by the red lines in Figure 4. However, as discussed in A. U. Abeysekara et al. (2021b) and V. Joshi (2019), there is a smearing effect caused by PSs not appearing point-like with this method.

To rectify this, a PS with HAWC J2031+415's index of 1.94 is simulated at PSR J2032+4127's location. This simulated source is then handled in the same method described above. The 1D Gaussians found with the simulated source are the "smearing" effect and are shown by the dashed blue line in Figure 4. This effect can now be subtracted out in quadrature with the observed data fits.

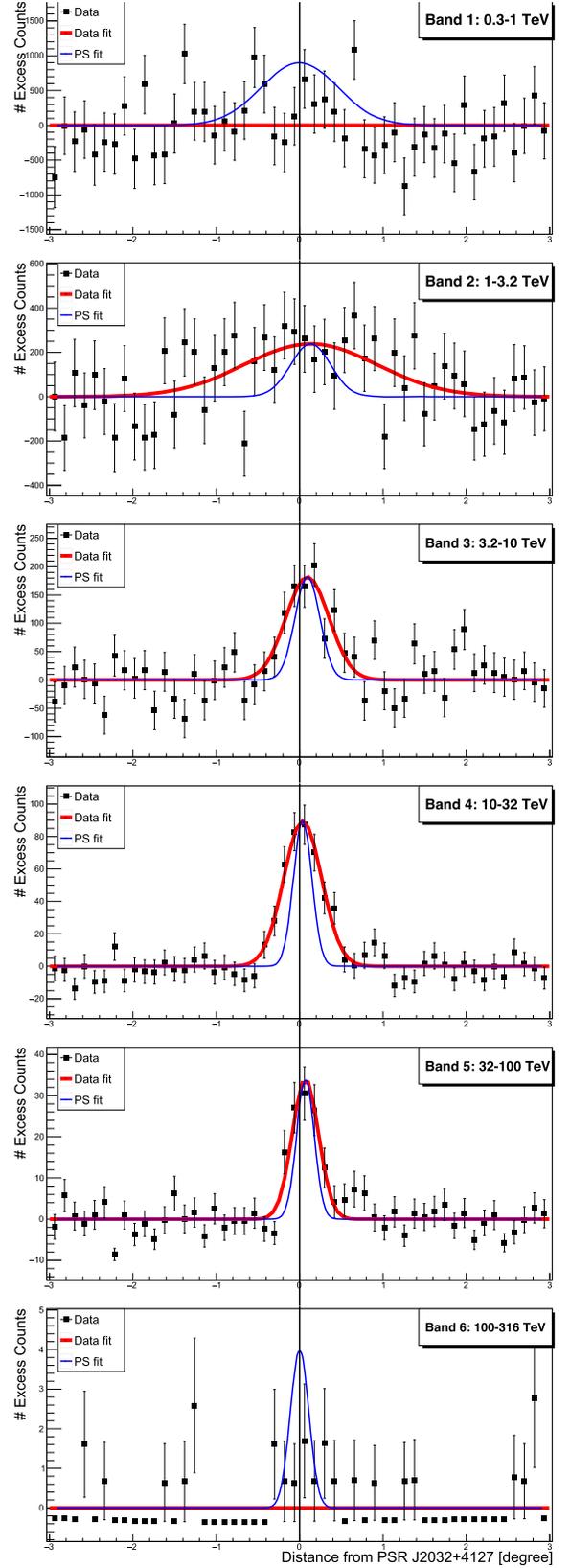

**Figure 4.** The longitudinal profiles for the excess count maps for HAWC J2031 +415. The red fitted lines correspond to the fitted Gaussians of each band while the blue dashed lines are the simulated PS Gaussians discussed in Section 4. The location of PSR J2032+4127 is indicated by the vertical line. The distance between HAWC J2031+415's best-fit centroid location and the pulsar's location is $0°.13$ or about 3 pc.

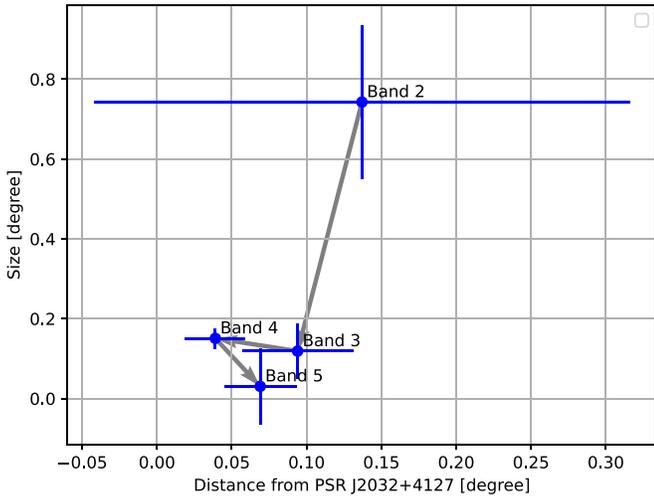

**Figure 5.** The results from the energy morphology study as described in Section 4. HAWC J2031+415's true size is presented on the y-axis and the distance to the pulsar is on the x-axis.

If a band is fit and a source cannot be identified, then it is considered a null result and a flat line is plotted. This indicates that either we cannot resolve the source from the background or the source is not visible to HAWC in that energy regime. Additionally, if the excess counts are negative, this means that, for the selected bin, there are more background events compared to data events.

### 4.2. Results

From Figure 4, there are no significant detections in bands 1 and 6; this is most likely caused by the spectrum of HAWC J2031+415. This source is not significantly detected in GeV or high TeV energies, which is what these two bands primarily comprise of. While there are fits for all other bands, band 2 requires more investigation. Its fit is diffuse and was checked against the diffuse background emission model to ensure the observed emission was from HAWC J2031+415 and not a large background fluctuation. The emission is observed at the $5\sigma$ level and is confirmed as a positive detection of HAWC J2031+415. Bands 3, 4, and 5 all show significant detections.

Figure 5 shows the size of TeV emission with increasing energy and its location with respect to PSR J2032+4127. The size is the true width of emission after subtracting both the PS smearing and any systematic offsets between data and simulation. This true width is given by

$$\sigma_{\text{true}} = \sqrt{\sigma_{\text{fit}}^2 - (\sigma_{\text{sim}} + \sigma_{\text{offset}})^2}. \quad (7)$$

While some faint energy-dependent morphology is present, particularly in the morphology shift from bands 2 to 3, there is no discernible trend at higher energies. Likewise, while there is a faint trend toward the pulsar's location at lower energies, there is nothing conclusive at higher energies. Additionally, this study was done on the best-fit location of HAWC J2031+415 to observe any potential shifting. The best-fit locations for each band trend toward the location in Table 1 and a slight though inconclusive change in morphology was detected.

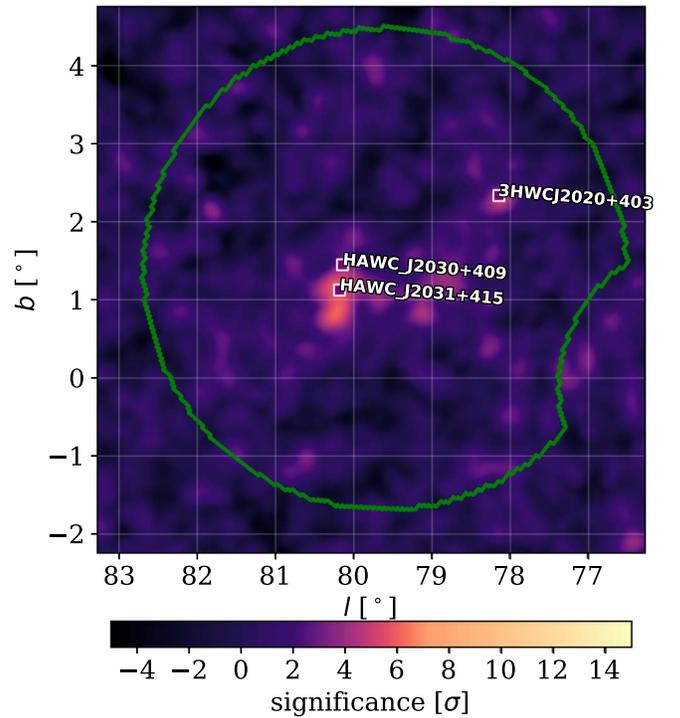

**Figure 6.** 172 days of HAWC data from 2017 May 26 to November 14. The maximum significance is located near HAWC J2031+415 and is $6.10\sigma$ for this interval.

### 5. Periastron of PSR J2032+4127

On 2017 November 13, PSR J2032+4127 completed its periastron of MT91 213. For approximately 170 days before periastron, $\gamma$-ray flaring was observed by both VERITAS and MAGIC (A. U. Abeysekara et al. 2018b) from the binary system. HAWC has been proved to observe flares on a much shorter timescale (A. Albert et al. 2022) and so we investigate this flaring. To do this, a 172 day section of HAWC data from 2017 May 26 to November 14 along with a smaller ROI of 3° was selected. The significance map is shown in Figure 6.

One hot spot appears in the data with a significance of $6.10\sigma$. This is compared to a separate map made before any flaring was detected. This map has 170 days and runs from 2016 September 30 to 2017 March 19. A similar hot spot of $6.37\sigma$ is found. Next, the source search method introduced in Section 3 is used on both these maps to determine both the number of sources and their fluxes to see if any flaring can be resolved. The results for both maps produce a single PS located near HAWC J2031+415. While the extension test for both sources was rejected with a $\Delta TS \approx 15$, the extension for both sources was $0°.35$, significantly larger than HAWC J2031+415's extension. This may be due to contamination from HAWC J2030+409.

It is clear that, while HAWC does resolve a source in the 170 day maps, we lack sufficient data to properly resolve any flaring from the periastron and instead only detect diffuse emission from the PWN. Potentially, future background-rejection algorithms may yield more concrete results but currently HAWC does not detect the periastron of PRS J2032+4127 and MT91 213.

## 6. Multiwavelength Fitting

### 6.1. NAIMA Framework

The NAIMA software is a nonthermal modeling framework that utilizes Markov Chain Monte Carlo calculations (V. Zabalza 2010). It has both leptonic and hadronic models that take flux points like the ones shown in Figure 2 as input and fit different emission models to them. At the γ-ray regime, the leptonic model considers IC scattering of relativistic electrons off low energy photons from the cosmic microwave background (CMB) radiation and far-infrared (FIR) and near-infrared (NIR), while the X-ray regime considers SYN emission released by high energy electrons moving through magnetic fields. The hadronic model considers $\pi^0$ decay (PD) from proton–proton collisions.

### 6.2. Methodology

We break the multiwavelength data introduced in Section 1 into two groups: TeV data fitted with the IC and PD models, and the lower energy range data that is fit with the SYN model. The former is the HAWC flux points presented in Figure 2 and serve as the high energy constraint for the models while the latter requires a more in-depth discussion.

The two X-ray observations from Suzaku (H. Murakami et al. 2011) and XMM-Newton (D. Horns et al. 2007) both detected diffuse X-ray emission inside the COG of HEGRA's and, by extension, HAWC's source extent. As discussed in H. Murakami et al. (2011), they were able to resolve two small (≃0°.01) distinct substructures, one of which they associate as the X-ray PWN produced by PSR J2032+4127 and produced a spectrum for it. In contrast, D. Horns et al. (2007) only detect diffuse emission and not the small structures found by Suzaku. Currently, it is not clear which data correspond to the X-ray PWN but, for this analysis, we consider the Suzaku structure 1 detection to be the probably X-ray PWN and the XMM-Newton diffuse detection to be the upper limit for the X-ray emission.

Similar to the XMM-Newton detection, the radio observation from the VLA (J. M. Paredes et al. 2006) found faint diffuse radio emission in HEGRA's COG but did not resolve any substructures associated with the PWN. While some small emission coincident with Suzaku's structure 1 was observed in Figure 1 of J. M. Paredes et al. (2006), no detailed analysis was presented on that emission. Therefore, as with XMM-Newton, the diffuse emission found by the VLA is considered as an upper limit of the radio emission produced by the PWN.

To model this data, four separate scenarios are considered: a leptonic model with IC handling the HAWC data and SYN modeling the lower energy data, a TeV only fit using IC and assuming different magnetic fields for the SYN model, a hadronic model only using PD to fit the HAWC data, and then another PD model with a fixed index at 2. This is to test whether there are enough TeV data to adequately constrain the hadronic index to the expected value of 2, assuming the primary acceleration mechanism comes from diffusive shock acceleration. The seed photon fields considered for the leptonic model are the CMB and values for the NIR and FIR approximated from C. C. Popescu et al. (2017). The column density needed to find the proton density is $7.7 \times 10^{21}$ cm$^{-2}$ from F. Camilo et al. (2009) which, considering HAWC J2031 +415's extent of ∼5.9 pc, gives a density of 417 protons cm$^{-3}$.

For both models, cutoff PL spectra are assumed for the proton and electron populations. This comes from observed flux points for HAWC J2031+415 and its best-fit spectral model. The parameters are the same as in Equation (4) with $E_p$ being set to 20 TeV for both models. Additionally, the energies required for the two models are given as $W_e$ for leptonic and both $W_e$ and $W_p$ for lepto-hadronic. The results from these models are shown in Figure 7 and Table 2.

### 6.3. Leptonic Results

From Figure 7, it is immediately clear that, while the TeV fits for the IC and the two PD models fit the HAWC data well, the magnetic field resulting from the combined IC and SYN field is quite low at 1.48 μG. Furthermore, applying different magnetic field values to the TeV only IC fit (the 0.5, 1, 3, and 6 μG lines) corroborates this low magnetic field. This leads to several different possible conclusions that are discussed below.

The first potential conclusion is that the X-ray source observed by Suzaku is the X-ray counterpart to HAWC J2031 +415. This would make the X-ray PWN approximately 13 times smaller than the TeV PWN. While significantly smaller than the TeV region, this may be expected for older systems. As discussed in S. P. Reynolds et al. (2012) and B. Olmi & N. Bucciantini (2023), X-ray bright PWNe require powerful young pulsars and, as this is an old system, it could explain the low energy and small size observed in X-rays. Such a low magnetic field may indicate that the SYN nebula is nearing the end of its life. In this case, more detailed diffuse radio measurements are needed to properly constrain the magnetic field. A more detailed discussion is presented in Section 5 of B. Olmi & N. Bucciantini (2023).

An alternative possibility is that the X-ray PWN is much more diffuse than Suzaku's approximately 0°.02 diffuse measurement. If we consider XMM-Newton's larger source and flux, then the fitted magnetic field increases to ≃3 μG, which, while still low, does approximately equal the Galactic average. While they are more diffuse X-ray emission observations, they could indicate that the X-ray PWN is significantly larger than Suzaku's detection and, by extension, give a larger average magnetic field.

One final consideration is the binary nature of this system. As discussed in B. Olmi & N. Bucciantini (2023), stellar winds from the companion star collide with the pulsar wind and create highly variable emission. It has been shown (A. A. Zdziarski et al. 2010) that X-ray emission may be produced by IC scattering off these dense photon fields and may significantly contribute to the observed X-ray emission. A detailed X-ray analysis would be needed to determine if this is occurring in the PSR J2032+4127–MT91 213 system but it could explain the small emission region and lower X-ray flux.

Another consideration is the energy budget $E_T$ of PSR J2032 +4127 and the relationship it has with the observed electron population. An approximation of $E_T$ can be found by multiplying $\dot{E} = 1.5 \times 10^{35}$ erg s$^{-1}$ by the pulsar's characteristic age of ∼200 kyr to give an energy budget of ∼$9 \times 10^{47}$ erg (see H.E.S.S. Collaboration et al. 2018). While the $E_T$ found is most probably lower than the actual energy budget, a comparison can still be drawn between $W_e$ and the observed $E_T$, which reveals that $W_e$ is ∼2% that of the energy budget for PSR J2032+4127. This value is reasonable (M. Di Mauro et al. 2019) and indicates that PSR J2032+4127 is capable of producing the observed electron population.

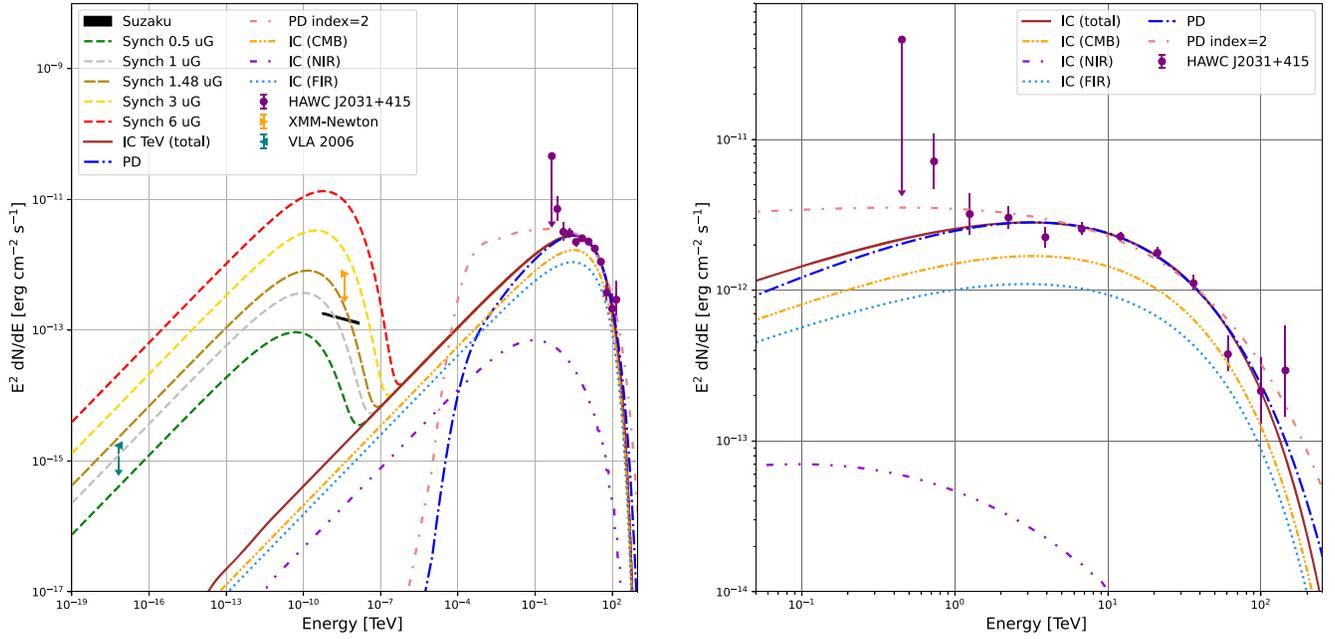

**Figure 7.** Left: the full multiwavelength fit results. The different SYN fit bands indicate the expected X-ray and radio fluxes for different magnetic fields from the IC fit using HAWC data. The 1.48 μG fit indicated by the thicker dashed line is the best-fit magnetic field using the Suzaku spectrum. Additionally, the IC HAWC data fit is broken into CMB, FIR, and NIR components. Right: zoomed-in TeV fit results. All three TeV models fit the data well, though it is clear that lower energy data are needed to properly constrain the PD model index.

**Table 2**
The Fit Values That Correspond to the Fitting Process Discussed in Section 6

| Model | $\log(\phi)(1/\mathrm{TeV})$ | $E_p$ (TeV) | $\alpha$ | $E_c$ (TeV) | $B\,(\mu\mathrm{G})$ | $W_{e,p} > 1$ TeV (erg) | AIC |
|---|---|---|---|---|---|---|---|
| IC + SYN | $42.8 \pm 0.1$ | 20 | $2.1 \pm 0.2$ | $59 + 13{-}22$ | $1.48 \pm 0.24$ | $(1.8 \pm 0.5) \times 10^{46}$ | 19.5 |
| IC only | $42.7 \pm 0.1$ | 20 | $2.1 \pm 0.3$ | $53 + 12{-}19$ | ⋯ | $(1.5 \pm 0.6) \times 10^{46}$ | 17.5 |
| PD free index | $43.1 \pm 0.06$ | 20 | $1.6 \pm 0.3$ | $139 + 36{-}52$ | ⋯ | $(3.1 \pm 0.5) \times 10^{46}$ | 17.9 |
| PD fixed index | $43.2 \pm 0.03$ | 20 | 2 (fixed) | $265 + 28{-}40$ | ⋯ | $(4.6 \pm 0.3) \times 10^{46}$ | 19.2 |

**Note.** As in Section 3, the parameters are the flux normalization, pivot energy, index, and the cutoff. The magnetic field for the SYN component is presented along with the separate electron and proton population energies. Finally, the AIC for all fits is presented. All uncertainties presented are statistical only.

### 6.4. Hadronic Scenario

For the hadronic models, both fit the HAWC data well visually and statistically. There are two key factors to note: the value of the free index fit compared to the fixed index fit and the found $E_c$ for both models. Both are discussed below.

The rather low value of the fitted index at $1.6 \pm 0.3$ compared to the expected value of 2 indicates that HAWC data alone may not be sufficient to properly constrain the hadronic index. It is clear from the fixed index fit that additional data, particularly in the GeV range, would greatly aid in constraining the index. There are GeV flux points from Fermi-LAT presented in A. U. Abeysekara et al. (2018a) that are matched to a 0°.14 extended source but they were obtained using a simplified model for the Cocoon and may contain contamination from the larger source. As such, they are excluded from this analysis.

Considering the $E_c$ for both fits along with the observed maximum energy of 151 TeV, would make HAWC J2031 +415 a potential PeVatron with both models having $E_c > 100$ TeV. Indeed, recent observations from the LHAASO observatory have detected >1 PeV events from this region (Z. Cao et al. 2021). However, it is extremely doubtful that the PWN is capable of producing such high energy events. A follow-up paper analyzing the PeV LHAASO sources (E. de Ona Wilhelmi et al. 2022) found that PSR J2032+4127's low spin-down luminosity is incapable of producing such high energy events. The binary nature of the PWN may help contribution to the hadronic model, but that is beyond the scope of this paper. It is clear that, though the HAWC data can be fitted well with a simple hadronic model, it is impossible that such a model is sufficient to power such high energy events.

### 6.5. Scenario Comparison

While both the leptonic and hadronic models fit HAWC's data well, there are concerns with both models. For the leptonic models, while the TeV data are fit well with the IC and IC + SYN models, the lack of significant X-ray and radio data makes it difficult to properly constrain the ambient magnetic field. The fitted value of 1.48 μG is significantly lower than the lower end of expected magnetic fields for PWNe (S. P. Reynolds et al. 2012) but this also could be a symptom of the X-ray

PWN fading away (B. Olmi & N. Bucciantini 2023). More radio data would help constrain the magnetic field of the PWN.

For the hadronic model, while it fits the HAWC data well, it is also physically impossible for the PWN to produce PeV protons given its low spin-down luminosity (E. de Ona Wilhelmi et al. 2022). Therefore, while it is included in this analysis for completeness, it is clearly not the preferred model for HAWC J2031+415's TeV emission.

## 7. Conclusions

The morphological studies we presented here reveal HAWC J2031+415 to be an extended emission region modeled as a symmetric Gaussian. As predicted by E. Aliu et al. (2014) and A. U. Abeysekara et al. (2018a), it has a spectral shape of a PL with exponential cutoff energy $E_c = 32$ TeV and is seen to 151 TeV. Given its close proximity to TeV J2032+4130, HAWC J2031+415 is most probably the high energy extension of this unidentified source. An energy-dependent morphology study found that, while there might be a downward trend in size, no significant conclusion was found. Additionally, a periastron study was done to observe whether HAWC saw any flaring, but none was detected.

We then performed a multiwavelength analysis considering radio and X-ray data using the NAIMA framework. We considered four models, two leptonic (one with lower energy data, one with only TeV data) and two hadronic (one with a fixed index and one with a free index). All four models fit the available data well, but the hadronic models are physically incompatible with the PWN scenario. The leptonic models found a very low magnetic field of 1.48 $\mu$G. This could indicate that the X-ray counterpart of the PWN may be fading, but more X-ray and radio data are needed to confirm this hypothesis. In conclusion, we present the highest energy observations of the PWN candidate HAWC J2031+415 and, by extension, TeV J2032+4130.

## Acknowledgments

We acknowledge the comments and suggestions from an anonymous referee.

We also acknowledge the support from the US National Science Foundation (NSF); the US Department of Energy Office of High-Energy Physics; the Laboratory Directed Research and Development (LDRD) program of Los Alamos National Laboratory; Consejo Nacional de Ciencia y Tecnología (CONACyT), México, grants 271051, 232656, 260378, 179588, 254964, 258865, 243290, 132197, A1-S-46288, A1-S-22784, CF-2023-I-645, LNC-2023-117, cátedras 873, 1563, 341, and 323, Red HAWC, México; DGAPA-UNAM grants IG101323, IN111716-3, IN111419, IA102019, IN106521, IN114924, IN110521, and IN102223; VIEP-BUAP; PIFI 2012 and 2013 and PROFOCIE 2014 and 2015; the University of Wisconsin Alumni Research Foundation; National Research Foundation of Korea (RS-2023-00280210); the Institute of Geophysics, Planetary Physics, and Signatures at Los Alamos National Laboratory; Polish Science Centre grant, DEC-2017/27/B/ST9/02272; Coordinación de la Investigación Científica de la Universidad Michoacana; Royal Society—Newton Advanced Fellowship 180385; Generalitat Valenciana, grant CIDEGENT/2018/034; the Program Management Unit for Human Resources & Institutional Development, Research, and Innovation, NXPO (grant No. B16F630069); Coordinación General Académica e Innovación (CGAI-UdeG), PRODEP-SEP UDG-CA-499; and Institute of Cosmic Ray Research (ICRR), University of Tokyo. H.F. acknowledges support by NASA under award number 80GSFC21M0002. We also acknowledge the significant contributions over many years of Stefan Westerhoff, Gaurang Yodh, and Arnulfo Zepeda Domínguez, all deceased members of the HAWC collaboration. Thanks to Scott Delay, Luciano Díaz, and Eduardo Murrieta for technical support.

## ORCID iDs

D. Avila Rojas https://orcid.org/0000-0002-4020-4142
H. A. Ayala Solares https://orcid.org/0000-0002-2084-5049
R. Babu https://orcid.org/0000-0002-5529-6780
E. Belmont-Moreno https://orcid.org/0000-0003-3207-105X
K. S. Caballero-Mora https://orcid.org/0000-0002-4042-3855
T. Capistrán https://orcid.org/0000-0003-2158-2292
A. Carramiñana https://orcid.org/0000-0002-8553-3302
S. Casanova https://orcid.org/0000-0002-6144-9122
U. Cotti https://orcid.org/0000-0002-7607-9582
J. Cotzomi https://orcid.org/0000-0002-1132-871X
S. Coutiño de León https://orcid.org/0000-0002-7747-754X
E. De la Fuente https://orcid.org/0000-0001-9643-4134
C. de León https://orcid.org/0000-0002-8528-9573
D. Depaoli https://orcid.org/0000-0002-2672-4141
N. Di Lalla https://orcid.org/0000-0002-7574-1298
B. L. Dingus https://orcid.org/0000-0001-8451-7450
M. A. DuVernois https://orcid.org/0000-0002-2987-9691
J. C. Díaz-Vélez https://orcid.org/0000-0002-0087-0693
K. Engel https://orcid.org/0000-0001-5737-1820
T. Ergin https://orcid.org/0000-0003-2423-4656
C. Espinoza https://orcid.org/0000-0001-7074-1726
K. L. Fan https://orcid.org/0000-0002-8246-4751
N. Fraija https://orcid.org/0000-0002-0173-6453
J. A. García-González https://orcid.org/0000-0002-4188-5584
M. M. González https://orcid.org/0000-0002-5209-5641
J. A. Goodman https://orcid.org/0000-0002-9790-1299
J. P. Harding https://orcid.org/0000-0001-9844-2648
I. Herzog https://orcid.org/0000-0001-5169-723X
D. Huang https://orcid.org/0000-0002-5447-1786
F. Hueyotl-Zahuantitla https://orcid.org/0000-0002-5527-7141
P. Hüntemeyer https://orcid.org/0000-0002-3302-7897
A. Iriarte https://orcid.org/0000-0001-5811-5167
J. Lee https://orcid.org/0000-0002-2153-1519
H. León Vargas https://orcid.org/0000-0001-5516-4975
A. L. Longinotti https://orcid.org/0000-0001-8825-3624
G. Luis-Raya https://orcid.org/0000-0003-2810-4867
K. Malone https://orcid.org/0000-0001-8088-400X
J. Martínez-Castro https://orcid.org/0000-0002-2824-3544
J. A. Matthews https://orcid.org/0000-0002-2610-863X
P. Miranda-Romagnoli https://orcid.org/0000-0002-8390-9011
E. Moreno https://orcid.org/0000-0002-1114-2640
M. Mostafá https://orcid.org/0000-0002-7675-4656
M. Najafi https://orcid.org/0000-0002-7576-4799
L. Nellen https://orcid.org/0000-0003-1059-8731
M. Newbold https://orcid.org/0000-0001-9428-7572
M. U. Nisa https://orcid.org/0000-0002-6859-3944

R. Noriega-Papaqui ◉ https://orcid.org/0000-0001-7099-108X
Y. Pérez Araujo ◉ https://orcid.org/0000-0002-8774-8147
E. G. Pérez-Pérez ◉ https://orcid.org/0000-0001-5998-4938
C. D. Rho ◉ https://orcid.org/0000-0002-6524-9769
D. Rosa-González ◉ https://orcid.org/0000-0003-1327-0838
E. Ruiz-Velasco ◉ https://orcid.org/0000-0001-6939-7825
A. Sandoval ◉ https://orcid.org/0000-0001-6079-2722
M. Schneider ◉ https://orcid.org/0000-0001-8644-4734
A. J. Smith ◉ https://orcid.org/0000-0002-1012-0431
R. W. Springer ◉ https://orcid.org/0000-0002-1492-0380
K. Tollefson ◉ https://orcid.org/0000-0001-9725-1479
I. Torres ◉ https://orcid.org/0000-0002-1689-3945
R. Torres-Escobedo ◉ https://orcid.org/0000-0002-7102-3352
R. Turner ◉ https://orcid.org/0000-0003-1068-6707
F. Ureña-Mena ◉ https://orcid.org/0000-0002-2748-2527
E. Varela ◉ https://orcid.org/0000-0003-0715-7513
L. Villaseñor ◉ https://orcid.org/0000-0001-6876-2800
X. Wang ◉ https://orcid.org/0000-0001-6798-353X
I. J. Watson ◉ https://orcid.org/0000-0003-2141-3413
S. Yun-Cárcamo ◉ https://orcid.org/0000-0002-9307-0133
H. Zhou ◉ https://orcid.org/0000-0003-0513-3841